\documentclass[preprint,journal]{vgtc}           
\onlineid{0}

\vgtccategory{Research}

\vgtcpapertype{application/design study}

\title{EventBox: A Novel Visual Encoding for Interactive Analysis of Temporal and Multivariate Attributes in Event Sequences}

\author{%
  \authororcid{Luis \ Montana}{0000-0002-0511-0466},
  \authororcid{Jessica \ Magallanes}{0000-0003-0022-2036},
  \authororcid{Miguel \ Juarez}{0000-0002-5128-0976},
  \authororcid{Suzanne \ Mason}{0000-0002-1701-0577}.
  \authororcid{Andrew \ Narracott}{0000-0002-3068-6192} \\
  Lindsey van Gemeren,
  \authororcid{Steven \ Wood}{0000-0002-5647-1990} and
  \authororcid{Maria-Cruz \ Villa-Uriol}{0000-0002-3345-539X}
}

\authorfooter{
\item
L.~Montana\textsuperscript{*}, J.~Magallanes, and M-C~Villa-Uriol \textsuperscript{**} are with Computer Science. E-mail: l.montanagonzalez\textsuperscript{*},m.villa-uriol\textsuperscript{**}@sheffield.ac.uk.
\item M.~Juarez is with the School of Mathematics and Statistics, HELSI.
\item A~Narracott is with the School of Medicine and Population Health,  Insigneo.
\item S.~Mason is with the Centre for Urgent Care Research, School of Health and Related Research.
\item All authors above are with University of Sheffield.
\item
L.~van Gemeren and S.~Wood are with Sheffield Teaching Hospitals NHS FT.
}

\abstract{The rapid growth and availability of event sequence data across domains requires effective analysis and exploration methods to facilitate decision-making. Visual analytics combines computational techniques with interactive visualizations, enabling the identification of patterns, anomalies, and attribute interactions. However, existing approaches frequently overlook the interplay between temporal and multivariate attributes. We introduce EventBox, a novel data representation and visual encoding approach for analyzing groups of events and their multivariate attributes. 
We have integrated EventBox into Sequen-C, a visual analytics system for the analysis of event sequences. To enable the agile creation of EventBoxes in Sequen-C, we have added user-driven transformations, including alignment, sorting, substitution and aggregation. To enhance analytical depth, we incorporate automatically generated statistical analyses, providing additional insight into the significance of attribute interactions. 
We evaluated our approach involving 21 participants (3 domain experts, 18 novice data analysts). We used the ICE-T framework to assess visualization value, user performance metrics completing a series of tasks, and interactive sessions with domain experts. We also present three case studies with real-world healthcare data demonstrating how EventBox and its integration into Sequen-C reveal meaningful patterns, anomalies, and insights. These results demonstrate that our work advances visual analytics by providing a flexible solution for exploring temporal and multivariate attributes in event sequences.
}


\teaser{
    \centering
    \includegraphics[width=\linewidth, alt={Four EventBox encodings.}]{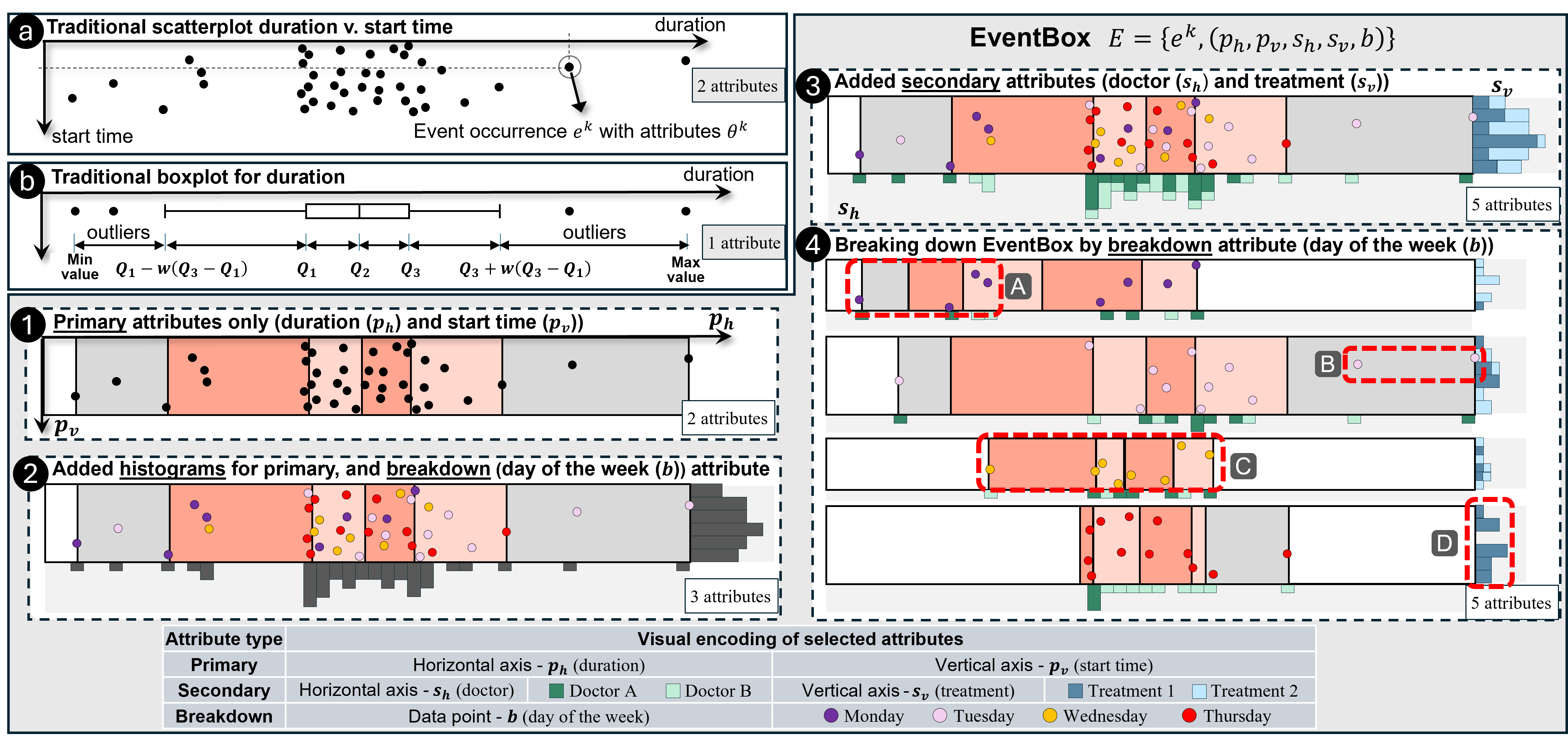}
    \caption{Four EventBox encodings (1-4) compared to conventional visualizations (a,b). (1-4) illustrate how a user would use multiple EventBox encodings to explore a selection of attributes. This sample dataset contains Monday to Thursday visits to a clinic for 40 patients undergoing two types of treatments offered by two doctors, with duration and appointment start times recorded. Findings using an EventBox breakdown are highlighted in red (A-D). 
    The table summarizes the selected attributes according to their role ($p$ - \textit{primary}, $s$ - \textit{secondary} or $b$ - \textit{breakdown}) and along which axis they are encoded  ($h$ - \textit{horizontal} or $v$ - \textit{vertical}).}
    \label{fig:teaser}
}


\graphicspath{{figs/}{Figures/}{pictures/}{images/}{./}} 

\usepackage{microtype}                 
\PassOptionsToPackage{warn}{textcomp}  
\usepackage{textcomp}                  
\usepackage{mathptmx}                  
\usepackage{times}                     
\usepackage{cite}                      
\usepackage{tabu}                      
\usepackage{booktabs}                  

\usepackage{multirow}
\usepackage{bm}
\usepackage{graphicx}
\usepackage{enumitem}
\usepackage{float}

\usepackage[utf8]{inputenc}
\usepackage{dirtytalk}
\usepackage{mathptmx}                  
\usepackage{listings}
\usepackage[normalem]{ulem}

\usepackage[switch]{lineno}

\def\sequenCP{Sequen-C~}

\def\mainPatternsE{RE1}  
\def\outlierPatternsE{RE2} 
\def\mainPatternsS{RS1}    
\def\outlierPatternsS{RS2}
\def\focusAnalysis{RS3} 
\def\transform{RS4}     
\def\eventboxE{RE3}  
\def\eventboxS{RS5}    
\def\stats{RE4}        
\def\quant{RS6}         

\begin{document}


\firstsection{Introduction}

\maketitle

The exponential growth of longitudinal event data in fields such as healthcare, finance, and social sciences has created a demand for advanced data exploration and analysis techniques, for example, to support data-driven decision-making.
Interactive visual data analytics has emerged as a user-centric approach to explore event sequences, which are ordered events occurring over time, typically characterized by timestamps and other multivariate attributes, including event type and duration. 

Existing techniques for visualizing event sequences primarily focus on encoding common pathways or sequential patterns, often overlooking the explicit representation of temporal and multivariate attributes \cite{guo2022survey,vanderlinden2023SurveyVisualization}. 
Building on our prior work \cite{magallanes2021sequen}, which introduced Sequen-C, an event sequence analysis system, and extending concepts in Magallanes \textit{et al.}~\cite{magallanes2019analyzing}, we address some of these challenges by making the following contributions:

\begin{itemize}[noitemsep]
  \item \textbf{EventBox:} A novel data representation and visual encoding technique for groups of events, simultaneously displaying up to five attributes, including temporal and multivariate attributes.  This facilitates the analysis of associations among multiple attributes, supported by a statistical report quantifying their significance.
  
  \item \textbf{Integration of EventBox into Sequen-C:} To support the agile creation of EventBoxes, we add a set of user-driven transformations to Sequen-C.
  
  \item \textbf{Three case studies:} Using real-world datasets, we demonstrate how Sequen-C and EventBox assist users in revealing patterns, anomalies and attribute relationships.
  
  \item \textbf{User evaluation:} To evaluate the effectiveness of our approach, we use the ICE-T framework \cite{wall2019Heuristic}, user performance metrics, and interactive sessions with experts.
  
\end{itemize}

\begin{figure*}[ht!]
   \includegraphics[width=\linewidth]{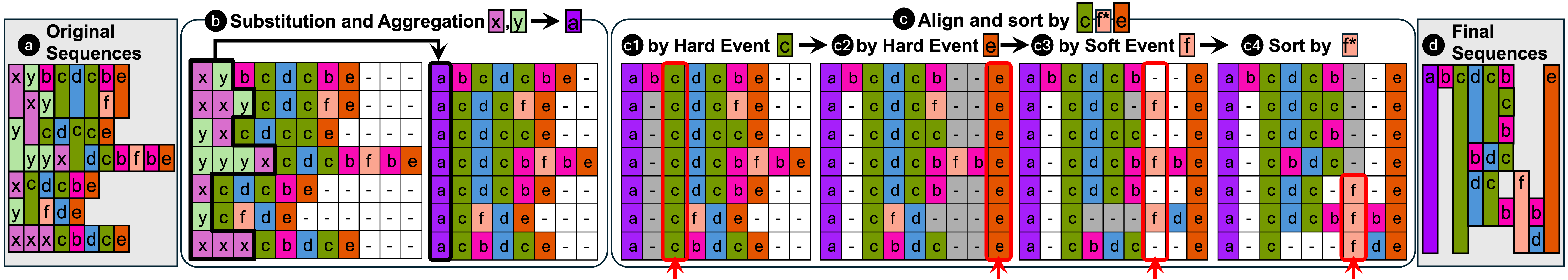}
   \caption{Example of concatenation of transformations. (a) Original sequences. (b) Events \includegraphics[height=0.4cm]{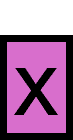} and \includegraphics[height=0.4cm]{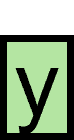} are replaced and aggregated by a new event type \includegraphics[width=0.2cm]{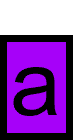}. (c1-4) Alignment and sorting by \includegraphics[height=0.4cm]{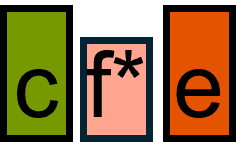}, where alignment by hard events \includegraphics[height=0.4cm]{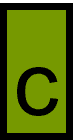} and \includegraphics[height=0.4cm]{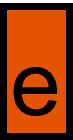} is applied first, then by soft event \includegraphics[width=0.2cm]{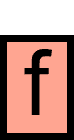}, and finally sorting by event \includegraphics[height=0.4cm]{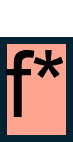}. (d) Final sequences.}
   \label{fig:Transformations}
\end{figure*}

Figures \ref{fig:teaser}, \ref{fig:Transformations} and \ref{fig:EventBoxCluster} illustrate various aspects of our work. Typically, a user interested in the analysis of the temporal event sequences representing the visits to a clinic would start with the overview shown in \autoref{fig:Transformations}. Based on the original sequences, the user would suggest aggregating and substituting events, as well aligning to concentrate on an event of interest, in this case the consultation with a doctor (event `f'), that could be further analyzed using the encoding shown in \autoref{fig:teaser}. \autoref{fig:teaser} presents four EventBoxes representing the same data through multiple attribute selections, comparing them to conventional visualizations such as (a) a scatter plot and (b) a boxplot. \autoref{fig:teaser}(1) illustrates the relationship between duration and start time, with colored bands showing quartile breakdowns and outliers. \autoref{fig:teaser}(2) adds histograms for duration and start time, providing further distribution details. The coloring by day of the week (\autoref{fig:teaser}(3)) highlights that shorter visits are more frequent on Mondays (A), while longer duration outliers occur on Tuesdays (B). This is confirmed by breaking down the EventBox by day of the week, as seen in \autoref{fig:teaser}(4). \autoref{fig:teaser}(4) reveals that there are no outliers on Wednesdays (C), and only Treatment 1 occurs on Thursdays (D). 
\autoref{fig:EventBoxCluster} represents another case, where the user would be interested in studying how time attributes (duration, start time and day of the week) relate to each other for a group of 9,003 event sequences that would represent the visits to a hospital.

\section{Related Work}\label{sect:related}
Despite progress in visual analytics for event sequence data, key challenges remain. These include handling data quality, uncertainty, scalability, heterogeneity, multivariate visualization, interpretability, and causality analysis \cite{guo2022survey}. Further difficulties involve comparing sequences with temporal and attribute-based complexity, defining similarity metrics, and managing granularity \cite{vanderlinden2023SurveyVisualization}. There is also a recognized need for improved interaction techniques and evaluation frameworks to reduce cognitive load and assess analytical effectiveness \cite{keim2011SolvingProblems}. 

Building on these, we introduce EventBox to address the \textit{explicit capture of temporal aspects of events} in the visualization \cite{vanderlinden2023SurveyVisualization}, the \textit{simultaneous visualization of multivariate attributes} \cite{guo2022survey,vanderlinden2023SurveyVisualization}, and the \textit{inclusion of meaningful similarity metrics} while improving their interpretability \cite{vanderlinden2023SurveyVisualization}. Integrating EventBox into Sequen-C has also required the use of \textit{interaction and interactive transformations} for the seamless navigation and effective exploration of the data \cite{keim2011SolvingProblems}.
This section reviews existing approaches in these four areas, highlighting their limitations and the gaps our work aims to address. We also discuss existing visualization evaluation strategies.

\subsection{Visual Analytics of Temporal Attributes}
In the analysis of event sequences, time attributes are vital. 
Several approaches represent temporal information visually, though the majority focuses on the ordering of events \cite{monroe2013temporal,perer2014frequence,wongsuphasawat2012exploring}, providing limited attention to the explicit summarization of time attributes. 
While sequential order conveys the time of occurrence of events, other time attributes, e.g.~time of the day or day of the week, are often not visually encoded in overviews. 
A variety of visual encodings have been explored to represent the temporal dimension. LifeLines2 \cite{wang2009temporal} employs histograms to show frequency distributions over time, allowing event analysis relative to a reference alignment point. However, sequences are not aggregated, and the frequency distribution is restricted to a selected event. TimeSpan \cite{loorak2016timespan} uses stacked bar charts for event durations in stroke treatments, while line charts facilitate analysis of trends, assuming a fixed event ordering. 

Other approaches encode duration by scaling the width of an event proportionally to the average duration, not displaying duration distributions or considering outliers.
Prior studies \cite{vrotsou2009activitree} indicate the need to identify infrequent sequences as outliers, and 
Van der Linden \textit{et al.}~\cite{vanderlinden2023SurveyVisualization}  highlighted that a key challenge in event visualization is handling events with unlikely durations, requiring tools to detect and appropriately represent such outliers.

Building on our earlier work \cite{magallanes2019analyzing}, we \textit{explicitly capture the temporal aspects of events} in the visualization of event sequences, allowing the encoding of time attributes, representing their distribution, and identifying outliers.

\subsection{Visual Analytics of Multivariate Attributes} 
Event sequences typically have associated a wide range of multivariate attributes. Currently, event sequence overviews represent them using average values or artificially created categories based on attribute values for events and sequences. Outflow \cite{wongsuphasawat2012exploring} and Frequence \cite{perer2014frequence} utilize a Sankey-inspired visualization approach in which edge hues correspond to average output values. These methodologies restrict their visual representation to mean values, not representing distributions. Di Bartolomeo \textit{et al.}~\cite{di2020sequence} offer an overview using a directed acyclic network, with sequentially arranged nodes, color-coded by attribute categories.
While this visualization indicates attribute value changes between events, it is limited to a singular attribute, lacking scalability with an increasing number of event types. EventPad \cite{cappers2018exploring} sorts sequential patterns according to attribute categories. These designs do not compare distributions of multiple attributes across sequential patterns. Linked views have also been used \cite{liu2009selltrend,borland2016multivariate}, including Treemaps, bar charts, and additional plots to visualize multivariate attributes. 

While these approaches provide valuable insights, they do not allow the \textit{simultaneous visualization and exploration of multivariate attributes within sequential patterns}. EventBox and its integration into Sequen-C supports both, explicitly linking the explored multivariate attributes to event sequences.  This approach enhances interpretability and facilitates the comparison of distributions across sequences.

\subsection{Visual Statistics} 
In statistics, high dimensional data visualization is often challenging to interpret and prone to misinterpretations \cite{padney, mcnutt, stein}. Identifying subtle relations within subgroups, 
requires many graphical representations, leading to a combinatorial increase in comparisons.  
Visual analytics facilitates pattern discovery and hypothesis generation, but like statistical methods, is susceptible to false patterns due to user-led data manipulation \cite{zhou2021Modeling}. While interactive tools aid exploration, incorporating quantities and statistical tests enhance interpretability and confidence in findings \cite{malik2016HighVolume}. 

Previous approaches \cite{malik2015cohort,malik2016HighVolume} use an overview, allowing cohort comparison, including summary statistics about the cohort, event sequences, time, event attributes, and sequence attributes. DecisionFlow \cite{gotz2014decisionflow} visualizes high-dimensional temporal event sequences, creating a directed graph of event sequences based on a user-defined query and calculating summary statistics, correlations, and odds ratios. CAVA \cite{zhang2015Iterative} visualizes patient groups and allows manipulation and creation of patient cohorts combining hierarchical and chart visualizations to display the summary statistics of a cohort. 
Cadence \cite{gotz2020VisualAnalysis} summarizes temporal event statistics and computes the correlation between an event occurrence and an outcome. Wentzel \textit{et al.}~\cite{wentzel2023DASS} explores radiotherapy cohort data to build predictive models for cancer patients using statistical bar charts that encode likelihood ratio test results to assess the correlation between clusters and the model outcome. 

While these techniques integrate statistical tests, they are predefined and lack flexibility in supporting diverse analytical needs. To increase user confidence in visual findings and facilitate data-driven decision-making, our work introduces the generation of user-driven automated statistical reports. These customized summaries enhance the \textit{interpretability of multivariate analysis of attributes in event sequences}. 

\subsection{Interaction}  
Interactive visualization is central to visual analytics \cite{tominski2020interactive}, enabling data exploration, insight generation, and feature extraction via operations such as selection, navigation, zoom, alignment, and visual comparisons.

Selection operations allow users to focus on data subsets for targeted analysis, and are often used in complex operations like filtering data or query building. Basic selection includes selection by attribute value \cite{jin2020CarePre,jin2021VisualCausality}, sub-sequences of events \cite{guo2019visual,chen2018StageMap}, interactive histogram bars \cite{xie2021VisualAnalytics,rogers2019Composer,jin2020CarePre}, and interactive sliders \cite{guo2018eventthread,du2020Interactive}. Building queries often involves the creation of visual queries \cite{monroe2012exploring,fails2006visual,du2017Finding,du2018VisualInterfaces}, regular expressions \cite{cappers2018exploring,squeries2015} and definition of milestone events \cite{gotz2014decisionflow}.

Navigation relies on interaction to alternate between visualization modes and layouts \cite{xu2019CloudDet}; zooming, offers a detailed view of a particular visualization segment \cite{guo2018eventthread,xu2019CloudDet}; and alignment, allows users to align temporal event sequences by selecting one \cite{wang2009temporal,wongsuphasawat2011lifeflow,monroe2013temporal,chen2018sequence} or multiple events \cite{polack2018Chronodes, magallanes2019analyzing}. 
The complexity of interactive visual comparisons varies depending on the application and user needs \cite{vanderlinden2023SurveyVisualization}.
Guo \textit{et al.}~\cite{guo2019VisualAnomaly} introduced interactive comparison glyphs designed to identify anomalous events within sequential data. Jin \textit{et al.}~\cite{jin2020CarePre} used histograms and event queries to select similar patients, and compare their medical history to aid clinical decision-making by predicting diagnostic events.

We have integrated \textit{interactive transformations into Sequen-C}, for seamless pattern exploration and for enabling the construction of EventBox representations for multivariate attribute analysis. 

\begin{figure*}[ht!]
   \includegraphics[width=\linewidth]{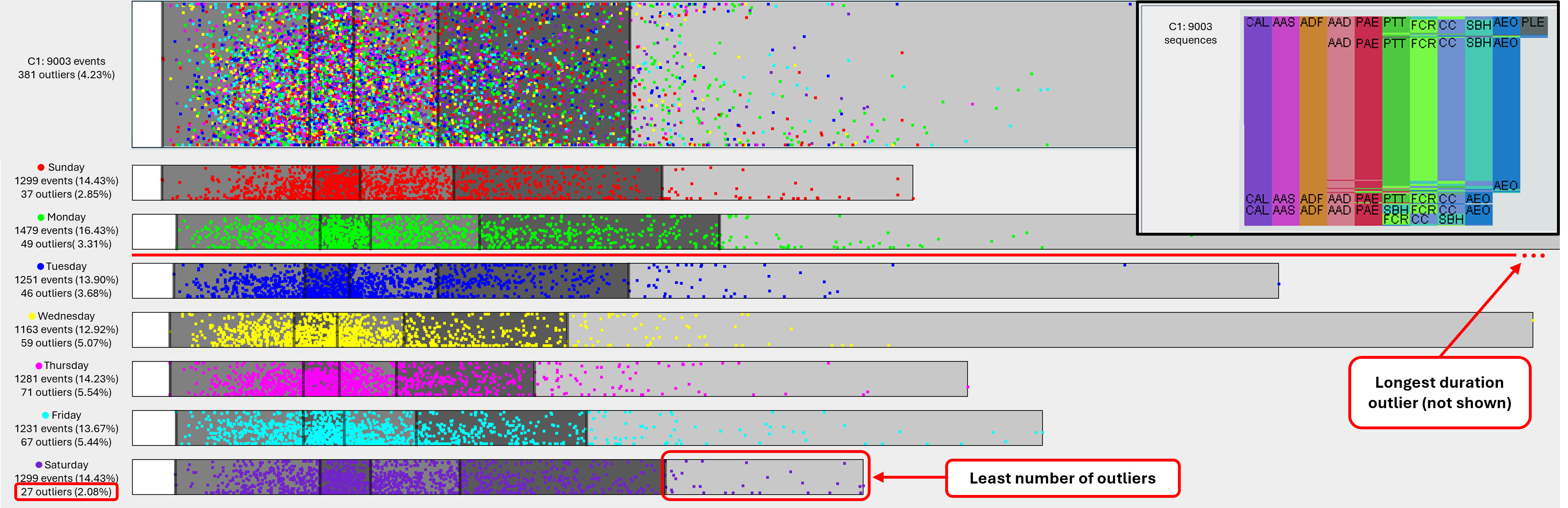}
   \caption{EventBox representation with breakdown by day of the week, after using a substitution and aggregation transformation to analyze 9,003 sequences. Findings are highlighted in red.}
   \label{fig:EventBoxCluster}
\end{figure*}
\subsection{Evaluation Approaches}  
Many evaluation methodologies assess the usability of visualization tools with questionnaires, interviews, or interactive sessions. Often, participants are asked to complete a series of tasks, and metrics such as accuracy and completion time measure the performance of the system under evaluation \cite{gotz2014decisionflow,gotz2016adaptive,wongsuphasawat2011lifeflow,xu2022exploring}. Other approaches include obtaining specialized feedback from domain experts after using the proposed visualizations \cite{malik2016HighVolume,guo2018eventthread,guo2019visual,gotz2020VisualAnalysis,jin2021VisualCausality, yeshchenko2022VisualDrift}. 

Typically, accuracy is prioritized over the value that visualizations can offer. To overcome this limitation, a value-driven evaluation methodology proposed \textit{ICE-T}, a questionnaire that assesses visualizations in terms of time minimization (T), insight generation (I), data essence conveyance (E) and data confidence generation (C) \cite{stasko2014Value,wall2019Heuristic}. 

In our work, to evaluate EventBox and Sequen-C, we use ICE-T to measure visualization value, accuracy and completion time to measure performance, and interactive sessions with domain experts alongside case studies with real-world datasets to validate domain applicability.

\section{Requirements Analysis}\label{sect:requirementsAnalysis}
The need for a novel dedicated strategy for analyzing temporal and multivariate attributes in event sequences emerged during the development of Sequen-C \cite{magallanes2021sequen,magallanes2019analyzing}. 
For example, domain experts needed solutions to reduce patient waiting times. To achieve this, they needed to understand how different attributes impacted those times.
To inform the design of EventBox, we conducted separate unstructured interviews with domain experts: a consultant in emergency services, and two IT hospital managers.
Each interview began with a walk-through of Sequen-C with a dataset relevant to their expertise.
We used open-ended questions to understand their analytical needs and challenges in depth. 
Based on the insights gained from these discussions and our prior experience analyzing temporal event sequences, we divide the requirements into two categories, those concerning EventBox and those required by the integration of EventBox into Sequen-C.
We also distinguish between two types of end users: \textit{domain experts}, who pose the questions, and \textit{expert analysts}, who perform the analyses. While domain experts may not always have specialized skills to conduct the analyses, we envision a typical use case involving collaboration, where the analyst operates the tool and the domain expert contributes domain-specific insight to guide and validate the findings.

\subsection{EventBox Requirements}
An EventBox encoding should allow the analysis of:
\begin{description}[align=left]
  \item [\mainPatternsE] \textbf{Main patterns:}  
Enable the detection and identification of the most frequent attribute values for a selection of data points.
  \item [\outlierPatternsE] \textbf{Outliers:} Enable the detection and identification of infrequent attribute values for a selection of data points.
  \item [\eventboxE] \textbf{Frequency distribution of multivariate attribute attributes:} Investigate relationships and frequency distribution of up to five attributes for a selection of data points.
  \item [\stats] \textbf{Statistical measures:} Incorporate quantitative summaries and, where possible, measures of statistical significance accompanied by an automatically generated report to support visual comparisons and mitigate bias.
\end{description}

\subsection{EventBox Integration into Sequen-C}
The integration of EventBox into Sequen-C should allow:
\begin{description}[align=left]

\item [\textbf{\mainPatternsS}] \textbf{Identification of main patterns:}  
Enable the detection and identification of the most frequent sequences, events, attributes, and attribute values.

\item [\textbf{\outlierPatternsS}] \textbf{Identification of outliers:} Enable the detection and identification of infrequent sequences, events, attributes, and attribute values.

\item [\focusAnalysis] \textbf{Focus and analysis:} Support the targeted analysis with details-on-demand of sequential patterns, subsets of sequences, event occurrences, and attribute values using interactive selection, preset filters, query-building, and coordinated views.

\item [\transform] \textbf{Interactive data transformations:} Facilitate comparisons via user-driven interactive data transformations, including event substitution, aggregation, alignment, and sorting.

\item [\textbf{\eventboxS}] \textbf{Flexible multivariate attribute analysis:} Support the interactive detailed analysis of multiple attributes. Attributes can be temporal (e.g., duration, start time), numerical (e.g., age), and categorical (e.g., gender, age group).

\item [\quant] \textbf{Quantitative summaries:} Incorporate quantitative summaries to support visual comparisons and mitigate bias.

\end{description}


\section{EventBox: Multivariate Analysis of Attributes}\label{section:methodology}
EventBox is a flexible data representation and visual encoding method designed to facilitate the simultaneous exploration of up to five multivariate attributes in event sequences. 
To analyze these attributes, occurrences of the same event type must be grouped. These occurrences share the same attributes at both the sequence level and the event level. For example, in \autoref{fig:teaser}, all grouped events include attributes such as duration, start time, day of the week, doctor, and treatment.
The grouping of events is supported by data transformations, such as aggregation, substitution, alignment, and sorting (see \autoref{section:transformations}).  
To illustrate its benefits and for its evaluation, we have integrated it into Sequen-C (see \autoref{section:system} and \autoref{section:evaluation}). 
\autoref{fig:teaser}(1-4) shows several EventBox configurations for the same dataset, highlighting that not all five attributes must be explored simultaneously. 

\subsection{Data Attributes Representation for Exploration}
We define an EventBox as the set $E = \{e^k, (p_h, p_v, s_h, s_v, b) \}$ where $k = 1, \dots, N$. $N$ denotes the number of time-stamped event occurrences in $E$. Each $e^k$ is an individual event occurrence, and $(p_h, p_v, s_h, s_v, b)$, the five attributes represented and explored in this EventBox. 
Each $e^k$ is associated with a collection of multivariate attributes $\mathbf\theta^k = (\tau^k, \kappa^k, \eta^k)$ that can be: temporal ($\tau$), categorical ($\kappa$), and numerical ($\eta$).
All events in an EventBox share a subset of attribute types represented by $\mathbf\theta$. They also share the same event type, which can result from an aggregation of multiple event types. Users will select $(p_h, p_v, s_h, s_v, b)$ from $\mathbf\theta$, designating two attributes as \textit{primary} ($p$: $p_h$ and $p_v$), two as \textit{secondary} ($s$: $s_h$ and $s_v$), and one as the \textit{breakdown} attribute ($b$), with subscripts $h$ and $v$ denoting the axis (horizontal or vertical) along which they are visually encoded.

In this paper, we focus on the analysis of temporal attributes. By default, we use duration ($p_h$) and start time ($p_v$) as primary attributes; however, other attributes could have been chosen (e.g. see \autoref{fig:4ClustersCURE}(c) where a categorical attribute is selected as $p_v$).

\subsection{Visual Encoding and EventBox Breakdown}
The visual encoding of an EventBox is inspired by box plots \cite{tukey1977boxplot}, scatter plots, and histograms (see \autoref{fig:teaser}(a,b)). 
EventBox uses four marks: container area, quartile lines, data points, and histograms. These marks utilize channels to encode the attributes in $\mathbf\theta$, the total number of events and the event type.

\textbf{Container area.} A rectangular box contains all $N$ event occurrences. \textbf{Color hue} encodes the event type, \textbf{height} is proportional to $N$, and \textbf{width} to the maximum value of the primary horizontal attribute ($p_h$) across all events (in \autoref{fig:teaser}, duration). 

\textbf{Quartile lines.} 
The distribution of the primary horizontal attribute ($p_h$) across all events is represented on the container's horizontal axis. The five statistics used in a traditional boxplot divide the container area: the minimum, 25\textsuperscript{th} percentile ($Q_1$), median ($Q_2$), 75\textsuperscript{th} percentile $(Q_3)$, and maximum values. These values are encoded using line marks and their \textbf{horizontal position} (see \autoref{fig:teaser}(1-4)). To improve visual clarity, alternating regions between quartile lines are shaded with varying \textbf{color saturation}, white is used for areas with no events and light grey when outliers are present.  
We use Tukey's definition to identify outliers as points outside the range $[Q_1-w(Q_3-Q1),Q_3+w(Q_3-Q1)]$ \cite{tukey1977boxplot}, where $w=1.5$ \cite{frigge1989some}.

\textbf{Data points.} \label{subsub:dataPoints}
Individual event occurrences $e^k$ are represented as point marks. Similar to scatter plots, their \textbf{horizontal position} encodes the primary horizontal attribute ($p_h$), and their \textbf{vertical position} the primary vertical attribute ($p_v$). The vertical axis is scaled top-to-bottom, covering the range of the $p_v$ attribute (in \autoref{fig:teaser}, start time). 
Points may encode additional information through \textbf{color hue} (e.g. the breakdown attribute ($b$)), or \textbf{transparency} to produce a heatmap representing the density of events.

\textbf{Histograms.} 
They support the exploration of primary and secondary attribute distributions. \autoref{fig:teaser}(2,3) illustrates how histogram bars transform into stacked bar charts after selecting secondary attributes, enabling their visualization in relation to primary attributes. Attribute values are \textbf{color-coded}, while bar \textbf{heights} indicate frequency (see \autoref{fig:teaser}(3,4)).

\textbf{EventBox break down.}
An EventBox can be subdivided based on the breakdown attribute ($b$), resulting in one EventBox per unique value of $b$. \autoref{fig:teaser}(4) demonstrates a breakdown by the day of the week, producing four EventBoxes corresponding to Monday, Tuesday, Wednesday and Thursday.

\subsection{Statistical Report}

In statistics, visualization and numerical summaries go hand in hand when teasing out relations among variables. 
We complement EventBox with an automatically generated report focusing on the user-selected attributes of interest (\textit{primary}, \textit{secondary} and \textit{breakdown}).  The report includes three statistical summaries. First, averages and standard deviations for continuous attributes, with mean comparison tests at different levels of user-defined granularity. 
Second, contingency tables analyze the associations of classifications based on the selected categorical attributes.
Lastly, the analysis of variance (ANOVA) tables relate a user-selected continuous attribute of interest to one or more categorical ones to help in model building via variable selection. The main distributional assumption underlying these tests holds thanks to the central limit theorem, due to the large group sizes \cite{schmider}, not requiring additional checks \cite{troncoso}.
\autoref{fig:4ClustersCURE}(d) shows a detail of the generated report relevant to the analyzed case study.

\begin{figure*}[t!]
    \centering
    \includegraphics[width=\textwidth]{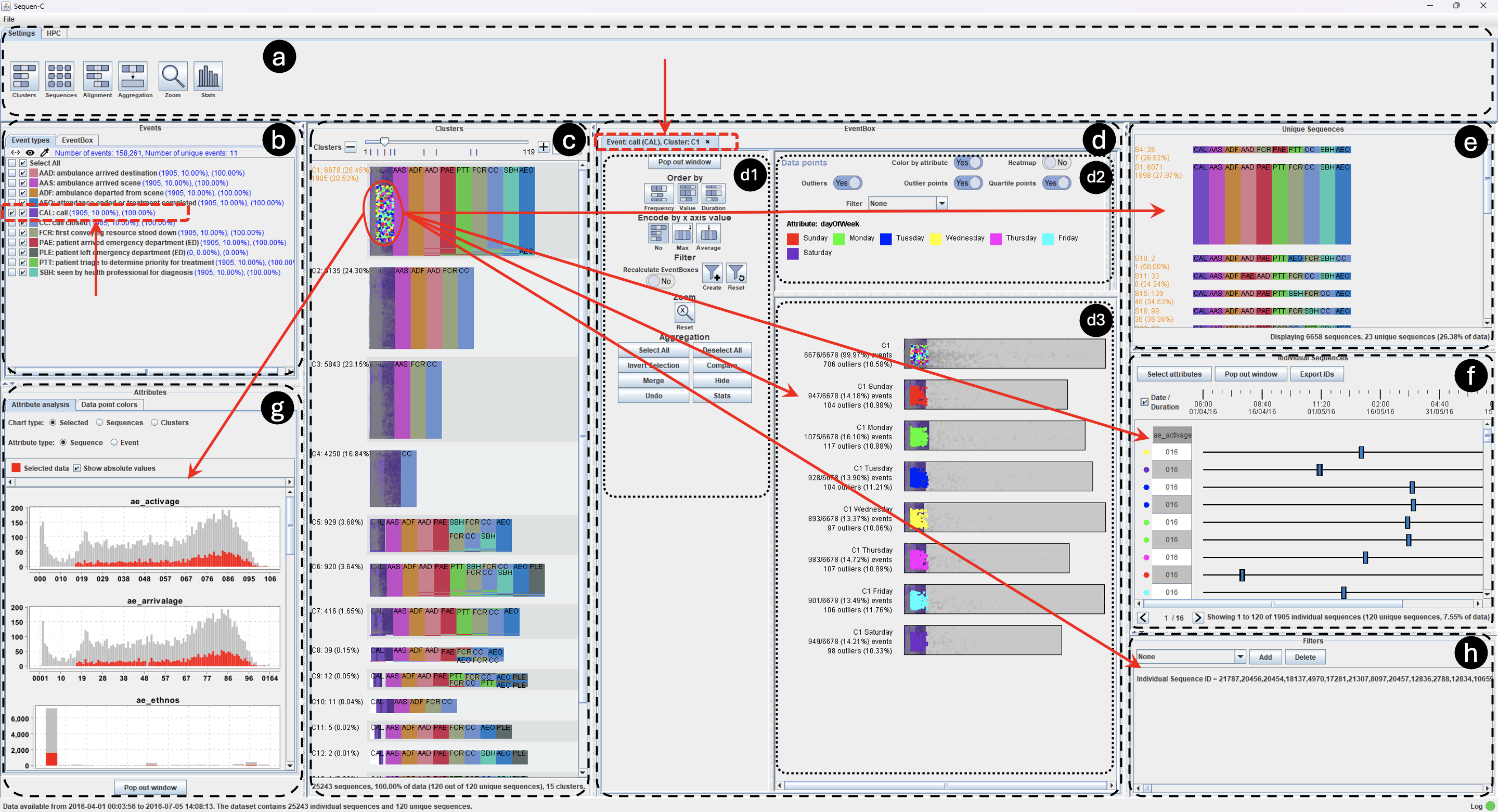}
    \caption{Sequen-C panels and coordinated views. (c) Clusters panel with 15 clusters. (b) In Events, \textit{CAL} (call) is selected and broken down by day of the week (\textit{breakdown} attribute) in (d3) EventBox (\textit{red dashed rectangles}). A subset of sequences in cluster C1 is selected (\textit{red circle}), updating (\textit{red arrows}) all panels (d-h). The controls in (a) provide access to substitution, aggregation and alignment transformations, among others.}
   \label{fig:systemOverview}
\end{figure*}


\section{Data Transformations} \label{section:transformations}
We use interactive data transformations to facilitate visual comparisons. Users can control how individual events are grouped and visualized by using substitution, aggregation, alignment, and sorting of event types. 
\autoref{fig:Transformations} illustrates how these operations work together to reveal how individual events relate to each other within groups of sequences. These transformations have been integrated into Sequen-C to identify events, groups of events or sequential patterns of interest that can undergo a deeper exploration of their attributes using EventBoxes.

\subsection{Substitution and Aggregation}
Users can substitute existing event types with newly created event types and aggregate the consecutive occurrences by merging their attributes.  For example, \autoref{fig:Transformations}(b) illustrates the event substitution and aggregation of events `x' and `y' by newly created event `a'. 
\autoref{fig:EventBoxCluster} illustrates how this transformation facilitates the study of the duration for 9,003 sequences (see the top right). All the events in all sequences have been aggregated into a single event type. An EventBox for the newly created event and its breakdown allow the study of how the duration of sequences varies with respect to the day of the week. Monday is the day of the week when outliers have the longest duration, whereas Saturday presents the least outliers.

\subsection{Alignment} \label{sub:alignment}
In addition to the existing automatic alignment present in Sequen-C, we have introduced interactive manual alignment to provide users with more control over which event types of interest to investigate further.  Event types of interest can be defined as \textbf{hard} or \textbf{soft}. The alignment of hard events is prioritized, and then we align by soft events between pairs of hard aligned events.  \autoref{fig:Transformations}(c1-c3) illustrates how sequences are padded to have uniform length and gaps are inserted to achieve the final alignment. While alignment preserves event order, it introduces visual gaps that may be misinterpreted as real time intervals.  These gaps are added to support alignment and visual comparison, and should not be interpreted as actual durations between events. For more details, see the Appendix (Algorithm 1).

\subsection{Sorting} We use sorting to facilitate comparisons. \autoref{fig:Transformations}(c4) shows how sorting by event `f' changes the visualization compared to \autoref{fig:Transformations}(c3). The sorting event index is used to create sub-sequences starting from that position to the end of the individual sequence. Then, the sorting is done by comparing these sub-sequences to determine the sequence order. For more details, see the Appendix (Algorithm 2).


\section{The System} \label{section:system}
We have integrated EventBox and the proposed transformations into the visual analytics system Sequen-C. \footnote{For more information please visit: \url{http://bit.ly/3IyEUI6}}
This allows the evaluation of EventBox in a system for the analysis of temporal event sequences and to facilitate the creation and interaction with EventBoxes to derive analytic findings. Sequen-C provides multilevel overviews that offer details on demand for event sequences. Overviews can range from high-level groupings of similar sequences organized into clusters to low-level details, such as individual event attributes.

\autoref{fig:systemOverview} shows the system layout, including: (a) general settings and controls, (b) events, (c) clusters, (d) EventBox, (e) unique sequences, (f) individual sequences, (g) attribute analysis, and (h) filters.  All panels are \textbf{coordinated}, and user interactions and selections in one panel are always propagated to the other ones. We will expand only on the changes relative to our earlier work \cite{magallanes2021sequen}.

\subsection{Events Panel} \label{sub:events}
This panel lists all event types in the dataset. Basic summary statistics \textbf{(\quant)} are displayed in blue, indicating the total number of events and unique events, with the proportion of each event type relative to the total number of events and the number of sequences in the dataset that contain each event (\textbf{\mainPatternsS}, \textbf{\outlierPatternsS}).

\textbf{User interaction.} Color hue encodes event types, checkboxes control event visibility, and which events should use the EventBox visual encoding. \autoref{fig:systemOverview}(b) illustrates an example where all events are visible, and the event type \textit{CAL} is selected and encoded as an EventBox (\autoref{fig:systemOverview}(c and d3)).

\subsection{Clusters, Unique Sequences, and Individual Sequences Views}
The clusters view in Sequen-C uses the EventBox visual encoding for as many events as specified by the user in the Events panel. \autoref{fig:systemOverview}(c) shows the EventBox encoding for event \textit{CAL} without outliers.

\textbf{User interaction.}  Data points in an EventBox can be selected using a mouse-based lasso operation. \autoref{fig:systemOverview}(c) illustrates the propagation of a lasso selection to the other panels (d3-h).
Such selections enable users to focus their analyses \textbf{(\focusAnalysis)}, for example on common patterns \textbf{(\mainPatternsS)} and outliers \textbf{(\outlierPatternsS)}.

\subsection{EventBox Panel} \label{sub:eventBox}
This panel enables users to explore and analyze EventBoxes in detail. \autoref{fig:systemOverview}(d) shows an EventBox for the user-selected event \textit{CAL} in cluster C1, with \textit{primary} attributes (duration and start time), no \textit{secondary} attributes, and \textit{breakdown} attribute (day of the week). The left panel (d1) allows the generation of statistical reports \textbf{(\stats)} and the merging \textbf{(\eventboxS)} of EventBoxes. The top panel (d2) provides the EventBox customization settings. These include the selection of \textit{primary}, \textit{secondary} and \textit{breakdown} attributes (only the breakdown attribute values are shown), with options to show/hide outliers, data points, and heatmaps.  The central panel (d3) displays the EventBox \textbf{(\mainPatternsE, \outlierPatternsE, \eventboxE, \stats)}.

\textbf{User interaction.}  Data points in an EventBox can be selected using a mouse-based lasso operation \textbf{(\focusAnalysis)}. Clicking on a histogram or stacked bar \textbf{(\focusAnalysis)} selects all the data points represented by that bar.

\subsection{Attribute Analysis Panel} \label{sub:attributes}
This panel allows exploring attributes \textbf{(\mainPatternsS, \outlierPatternsS)} for groups of sequences \textbf{(\focusAnalysis)}.
Attributes can be analyzed at either the event level or the sequence level. \autoref{fig:systemOverview}(g) visualizes how the attributes of the selected sequences compare to those of the entire dataset through stacked bar charts, which can display absolute or relative values \textbf{(\quant)}.

\textbf{User interaction.} Users can filter or select sequences and attribute values within the charts. This supports detailed comparisons and focused analyses  \textbf{(\focusAnalysis)} of selected sequences.

\subsection{Filters Panel} \label{sub:filters}
Sequen-C heavily relies on user interaction to define meaningful groups of sequences for joint analysis. Users can achieve this by selecting data points in an EventBox; interacting with bar charts in an EventBox, and the Attribute analysis panel; or by selecting clusters, unique and individual sequences in the Clusters, Unique and Individual sequences panels, respectively.

\textbf{User interaction.} User selections are translated into structured queries (see \autoref{fig:systemOverview}(h)) \textbf{(\focusAnalysis)}. Users can also specify complex queries through a guided graphical user interface \textbf{(\focusAnalysis)}. For example, a query that selects individual sequences within cluster C1 and filters patients older than 50 years would be defined as: ``(Cluster ID = C1) AND (age $>$ 50)''. Sequences that meet the query criteria are selected and propagated across all the coordinated views.


\section{User Evaluation}\label{section:evaluation}
Twenty-one participants (3 domain experts, 18 novice data analysts) evaluated Sequen-C and EventBox. We evaluated EventBox and Sequen-C together, as we were interested in assessing the value of the EventBox visual encoding in the context of the analysis of temporal event sequences. The novice data analysts were engineers with experience in data analysis but no prior exposure to Sequen-C or the EventBox visual encoding, while domain experts had previous familiarity with the system, but not with all its functionalities. 
During the evaluation, novice data analysts were assigned specific tasks and performed the data analysis using Sequen-C. Separately, domain experts formulated questions, and an expert data analyst conducted the analyses on their behalf. The evaluation was designed and conducted by two researchers with expertise in Sequen-C and EventBox. No time constraints were imposed on participants.

Our study was ethically approved via the University of Sheffield’s Ethics Review Procedure. The Appendix (Section 2) includes all the questionnaires, mapping of questions to design requirements, and detailed user evaluation statistics.

\subsection{Study Design for Novice Data Analysts}
The novice data analyst evaluation (age range: 22-40, 9 female, 9 male) was conducted in person and included: training, user performance, and user evaluation.
Screens and mouse clicks were recorded to verify timings and analyze usage patterns.

\textbf{Training.} The researchers leading the evaluation gave a hands-on introduction to Sequen-C and EventBox. At the end, participants completed a questionnaire to verify their correct understanding and adequate use of Sequen-C and EventBox features.

\textbf{User performance.} Participants were given a 15-question multiple-choice questionnaire to assess \textit{accuracy} (correct answers ratio) and \textit{completion time} (per question response time). They were able to complete this part in less than 60 minutes. Questions were inspired by domain experts’ real-world needs and were designed to cover all design requirements. The first part of the questionnaire was designed to evaluate Sequen-C's interface and the interactive features required to create and customize EventBoxes. The remaining questions were designed to evaluate EventBox’s effectiveness. In particular, questions Q1–Q7 did not require EventBox, Q8 and Q9 could be answered without it but with significantly greater effort, while Q10–Q15 required EventBox. 

The expert analysts leading the evaluation assisted participants without revealing answers. 

\textbf{Visualization value and user feedback.} Participants completed the 21-question ICE-T questionnaire \cite{wall2019Heuristic} rating the system’s \textit{visualization value}, followed by free-text feedback on usability, strengths, and areas for improvement.

\subsection{Novice Data Analysts Results}
\textbf{Accuracy and Completion time.} Participants achieved an average accuracy of 90.37\% ($\pm10.03\%$). Question 3 (Q3) was the most challenging, with only 8 of 18 participants (44.44\%) answering correctly; all answer choices were plausible if a crucial step was skipped. The average response time per question was 95 seconds, ranging from 27 seconds (Q7) to 221 seconds (Q9). Despite requiring multiple steps and panel interactions, Q9 was answered correctly by all participants. Full results are in the Appendix (Tables 1 and 2).

\textbf{ICE-T questionnaire.} 
In ICE-T, scores range from 1 (strongly disagree) to 7 (strongly agree), with scores $\geq 5$ indicating strengths and $\leq 4$ weaknesses, and a recommended overall average of at least 5 for effective visualizations. Overall, for Visualization \textbf{Value} our system achieved $5.82\pm 0.53$.
The highest-rated component was \textbf{Insight} (6.04), highlighting the system's ability to facilitate intentional and incidental insights, suggesting that Sequen-C's overview and multi-level navigation enhance the identification of key patterns (\textbf{\mainPatternsS}, \textbf{\mainPatternsE}) and anomalies (\textbf{\outlierPatternsS}, \textbf{\outlierPatternsE}).
The \textbf{Time} component, assessing search efficiency, scored the lowest (5.7), indicating room for improvement in speed (more details in \autoref{section:discussion}).
Finally, for \textbf{Essence}, the highest rated heuristic (6) was ``The visualization helps understand how variables relate in order to accomplish different analytic tasks'', which highlights the value that the EventBox brings to study multivariate attributes (\textbf{\eventboxE}).
For \textbf{Confidence}, the lowest rated heuristic (5.06) highlighted the need for more explicit communication of data inconsistencies such as unexpected, duplicate, missing, or invalid data.
Full results are available in the Appendix (Table 3).

\textbf{User feedback.}
Participants provided valuable feedback on the visualization system. They appreciated its ability to quickly reveal patterns and answer questions efficiently, particularly highlighting the hierarchical organization of visualizations and the accessibility of descriptive statistics. However, they noted areas for improvement, such as the difficulty in selecting variables, a steep learning curve for beginners, and the need for improved responsiveness and usability. Suggestions for additional features included customizable layouts, enhanced discoverability of existing functions, and more options for analyzing subsequences.

\begin{figure*}[ht]
   \centering
   \includegraphics[width=.9\textwidth]{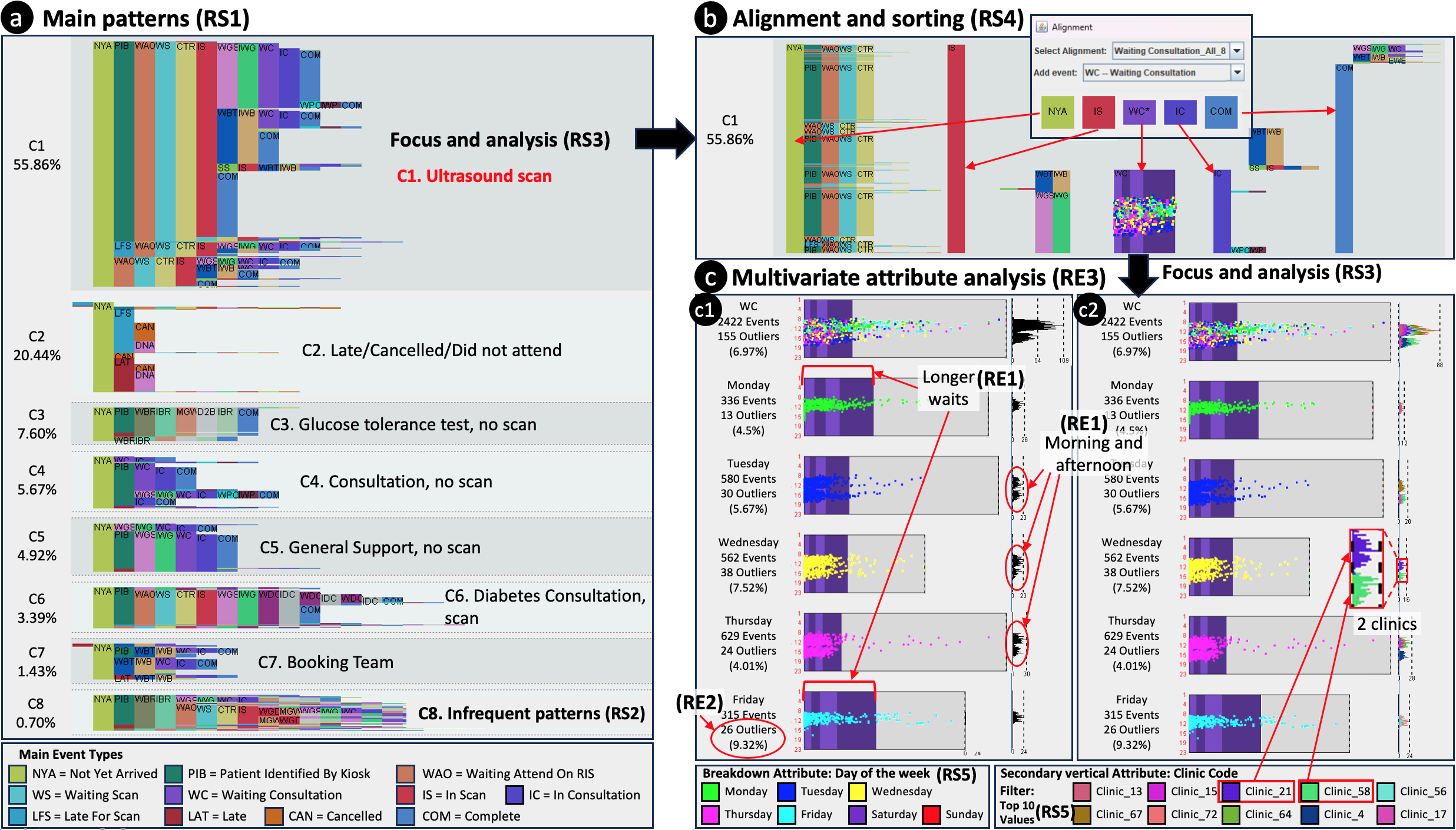}
   \caption{ ANCU case study. (a) 8 clusters overview showing main sequential patterns. (b) Alignment by \textit{NYA}, \textit{IS}, \textit{WC}, \textit{IC} and \textit{COM} events; and sorting by \textit{WC}. (c1) EventBox for \textit{WC} event displays the distribution of event occurrences by duration and start time (\textit{primary} attributes), with breakdown by \textit{day of the week} (\textit{breakdown} attribute). (c2) \textit{ClinicCode} attribute (top 10 most frequent values) is added as \textit{secondary} attribute (vertical axis). }
   \label{fig:Antenatal}
\end{figure*}

\begin{figure*}[ht]
   \centering
   \includegraphics[width=.9\textwidth]{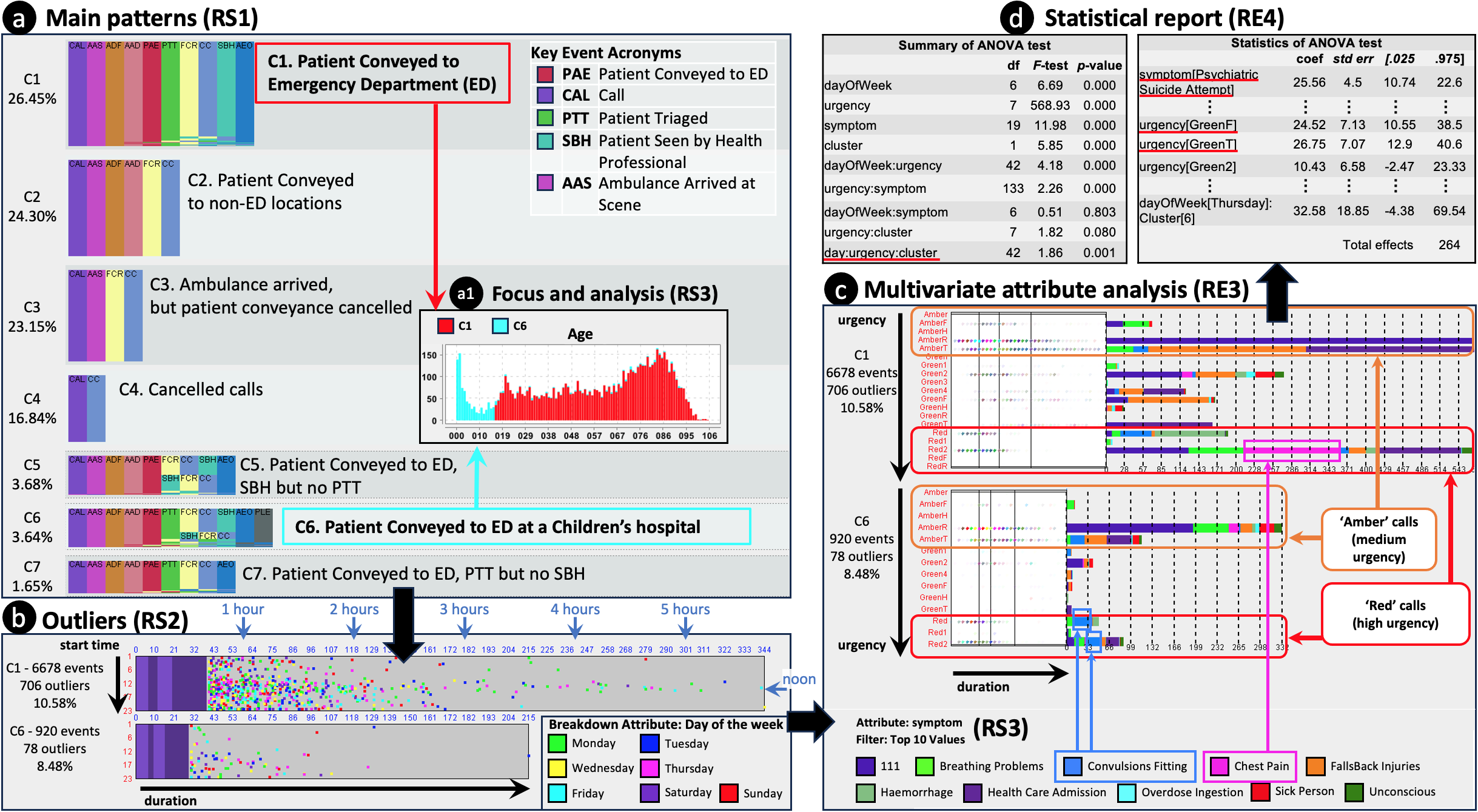}
   \caption{CUREd case study. (a) Main patient pathways overview. (a1) Attribute analysis for age shows differences between clusters C1 (adults) and C6 (minors). (b) EventBox shows quartiles (color bands) and outliers (data points) for the event call (\textit{CAL}) in C1 and C6. 
   (c) EventBox (as heatmap) for \textit{CAL} with \textit{primary} attributes (duration and urgency), and as \textit{secondary} attribute (reported symptom). 
   (d) Automated statistical report identifies attributes (\textit{left}) and specific attribute values (\textit{right}) impacting the duration of \textit{CAL}, highlighting the significant ones.}
   \label{fig:4ClustersCURE}
\end{figure*}

\subsection{Domain Experts Feedback}
We held separate interactive sessions with 3 domain experts in healthcare delivery (\textbf{E1}, \textbf{E2}, \textbf{E3}). Each session was divided into three parts, lasting 90 minutes. As they had previous experience with Sequen-C, an expert analyst provided a walkthrough of the new features of EventBox and Sequen-C in the first part. In the second part, the domain experts formulated questions, and the expert analyst answered them using EventBox and Sequen-C. In the last part, we asked the domain experts to give us feedback for each ICE-T category, general usability, and areas of future work. At the end, we asked them to put in writing their feedback.  This section presents their feedback complemented by our notes. 

\textbf{Insight}. Domain experts agreed that our visualization helps to identify patterns in complex processes quickly. \textbf{E2} noted its usefulness in detecting outliers and deviations. \textbf{E1} and \textbf{E2} found the EventBox breakdown valuable for identifying bottlenecks and areas for improvement, with \textbf{E1} highlighting its flexibility in achieving results through multiple approaches. \textbf{E2} praised the integration of charts and selection propagation for enhanced analysis. \textbf{E3}  emphasized the system's ability to reveal relationships between sequence patterns and their attributes.

\textbf{Time}. All domain experts found the system responsive and intuitive to get quick answers. \textbf{E2} highlighted its capacity for rapid data exploration and filtering without the need for complex SQL queries. \textbf{E2} believed that this feature enables users to utilize the system without prior knowledge of SQL, significantly broadening its accessibility. \textbf{E3} mentioned that the data can be quickly explored without a steep learning curve due to its multilevel overview and coordinated views. 

\textbf{Essence}. All domain experts agreed that the visualizations provide multiple perspectives of the data, going beyond the individual sequence.

\textbf{Confidence}. 
\textbf{E3} found that statistical analysis enhanced confidence in the findings and that multivariate attribute analysis helped validate known process timings. 
This leads us to think that confidence in the visualization depends on the user, as domain experts are reassured when familiar patterns are confirmed, increasing trust in the system. \textbf{E1} acknowledged that data accuracy is a common challenge in healthcare data, so the inclusion of strategies to help analysts with data quality assurance is essential.

\textbf{Usability}. \textbf{E2} thought that the user experience is well-suited for clinical data analysts, noting that while the system's basic features are straightforward and quick to learn, mastering its more complex functionalities demands additional time. However, \textbf{E2} emphasized that once these advanced features are learned, they offer substantial value, particularly when compared to traditional statistical analysis tools employed in the healthcare sector.

\textbf{Aspects to explore further}. Domain experts suggested additional features and applications for Sequen-C. \textbf{E1} proposed integrating a map with deprivation index overlays to analyze correlations with emergency department calls. \textbf{E3} noted that the selected clusters aligned with clinical pathways in their clinic and recommended exploring event aggregation before clustering to reduce unique sequences and create more meaningful clusters.


\section{Case Studies}\label{section:case_studies}

To demonstrate the value of EventBox and its integration into Sequen-C, we present three case studies using real-world medical datasets. 

\subsection{ANCU: Antenatal Care Unit}
This study, conducted in collaboration with experts from Sheffield Teaching Hospitals NHS Foundation Trust (United Kingdom), analyzes 73,279 recorded events from the Antenatal Care Unit outpatient clinic, tracking the visits of 9,623 pregnant women over three months, including consultations, ultrasound scans, and blood tests.

The analyst imported the dataset into Sequen-C and, after an initial inspection, selected 8 clusters for further exploration (\autoref{fig:Antenatal}(a)) \textbf{(\mainPatternsS)}. 

We chose to study patients waiting times for consultation of patients undergoing an ultrasound scan.  These patients were in cluster C1 \textbf{(\focusAnalysis)}. First, we aligned \textbf{(\transform)} by the sub-sequence \textit{NYA-IS-WC-IC-COM} (\textit{NYA} - Not Yet Arrived, \textit{IS} - In Scan, \textit{WC} - Waiting for consultation, \textit{IC} - In consultation, \textit{COM} - Visit completed). \textit{WC} and \textit{IC} were defined as soft events to ensure that they were between \textit{IS} and \textit{COM} events. And then, we sorted \textbf{(\transform)} by the \textit{WC} event to group all sequences containing it (\autoref{fig:Antenatal}(b)). To analyze the distribution of duration and starting time (\textit{primary} attributes), we built an EventBox for the \textit{WC} event in C1 (\autoref{fig:Antenatal}(b,c)).

To understand how waiting times varied during the week, we used as EventBox \textit{breakdown} attribute \textit{day of the week} \textbf{(\eventboxE)}. An initial analysis indicates that the \textit{WC} event on Fridays has a higher proportion of outliers compared to other weekdays \textbf{(\outlierPatternsE, \stats)}; and consultations on Mondays and Fridays are scheduled mainly in the morning with longer waiting times (duration), whereas Tuesdays through Thursdays have morning and afternoon sessions involving shorter waits (\autoref{fig:Antenatal}(c1)). To further investigate the variance in start times, we incorporated the \textit{ClinicCode} attribute into the vertical axis histogram (\textit{secondary} attribute), only including the top 10 most frequent values \textbf{(\focusAnalysis)}. This unveiled that different time slots are associated with distinct clinics. For example, Wednesdays show that morning consultations occur in \textit{Clinic\_21}, while afternoon sessions are linked to \textit{Clinic\_58} (\autoref{fig:Antenatal}(c2)). 

\subsection{CUREd: Ambulance Service Calls}
In collaboration with an expert from the Centre for Urgent and Emergency Care Research (CURE), we studied three months of data from the CUREd research database \cite{cureddata} containing 25,243 calls for 21,805 patients, and 34 attributes. CUREd compiles time-stamped events and demographics related to emergency service phone calls (999 or 111) within the Yorkshire and the Humber region. Calls result in ambulance conveyance to Emergency Departments (ED) or admissions to inpatient facilities, among others. 

The analyst imported the dataset into Sequen-C and selected 15 clusters. \autoref{fig:4ClustersCURE}(a) shows clusters C1-7 as the most representative \textbf{(\mainPatternsS)}. 
The expert was interested in studying how call pathways related to patient characteristics.

We chose to study the differences between clusters C1 and C6 as they exhibited a similar sequential pattern \textbf{(\mainPatternsS)}. 
\autoref{fig:4ClustersCURE}(a1) shows a snapshot of the attribute analysis panel displaying the age distribution for both clusters \textbf{(\focusAnalysis)}, revealing that C6 predominantly includes children. This insight prompted further analysis to compare both clusters' call durations and characteristics. To achieve this, the event type representing a call (\textit{CAL}) was used for alignment \textbf{(\transform)} and chosen for further analysis in an EventBox \textbf{(\eventboxE)} using the default \textit{primary} attributes (duration and start time). The duration of the event call includes the duration of the call itself and the wait until an ambulance is dispatched to the patient, hence the extremely long duration. \autoref{fig:4ClustersCURE}(b) illustrates this for cluster C1 \textbf{(\outlierPatternsE)},  displaying a notably long duration outlier (over 5 hours) and a higher proportion of outliers compared to cluster C6 (10.58\% vs. 8.48\%) \textbf{(\stats)}. An investigation into long-duration outliers indicated that they were mostly related to calls involving non-urgent patient transfer requests between a healthcare facility and the emergency department. Both clusters had a larger proportion of outliers in the afternoon \textbf{(\eventboxE)}, this could also be confirmed by the use of a histogram for start time (\textit{primary vertical} attribute), suggesting the cumulative effect of this type of request during the day. The EventBox colored bands show that in 75\% of calls, an ambulance arrived within 40 minutes for C1 and 30 minutes for C6 from the call's start.

The analyst examined the correlation between call duration (\textit{primary horizontal} attribute) and other factors, such as urgency (\textit{primary vertical} attribute) and symptom (\textit{secondary vertical} attribute), using an EventBox heatmap \textbf{(\eventboxE)}. \autoref{fig:4ClustersCURE}(c) shows that most of the ‘Red’ emergency calls (high urgency) in both clusters were dispatched quickly (within 10 minutes), whereas ‘Amber’ calls (medium urgency) had a more varied duration. This pattern could be also confirmed by breaking down by urgency and using a histogram for call duration. To analyze the distribution of symptoms relative to urgency, a histogram filtered to the 10 most frequent symptoms \textbf{(\focusAnalysis)} revealed that ‘Chest Pain’ was most common for ‘Red’ calls in cluster C1, while ‘Convulsions Fitting’ dominated in cluster C6.

\autoref{fig:4ClustersCURE}(d) presents a summary table and excerpts from a statistics table of an ANOVA test \textbf{(\stats)} from the automatically generated statistical report, analyzing call duration across selected categorical attributes (`day of the week', `urgency', `symptom', and `cluster number'). The summary table (\textit{left}) evaluates mean differences introduced by these attributes and their interactions, with results interpreted from bottom to top. The three-way interaction among ‘cluster’, ‘day of the week,’ and ‘urgency’ (last row) is significant (small $p$-value), indicating that at least one subgroup has a distinct average duration. Consequently, all two-way and main effects should be retained, even if their $p$-values suggest otherwise.
From the statistics table (\textit{right}), out of the 264 main effects and interactions, the 3 with the largest statistical significance increasing the call time are `symptom': `Psychiatric Suicide Attempt', and for `urgency': `GreenT' and `GreenF'. This suggests that calls categorized with those two codes are considered less urgent and might require a longer call to address. This is also consistent with phone calls related to a psychiatric suicide attempt, where the operators would keep the caller engaged longer in the call.

\begin{figure}[t]
   \centering
   \includegraphics[width=0.9\linewidth]{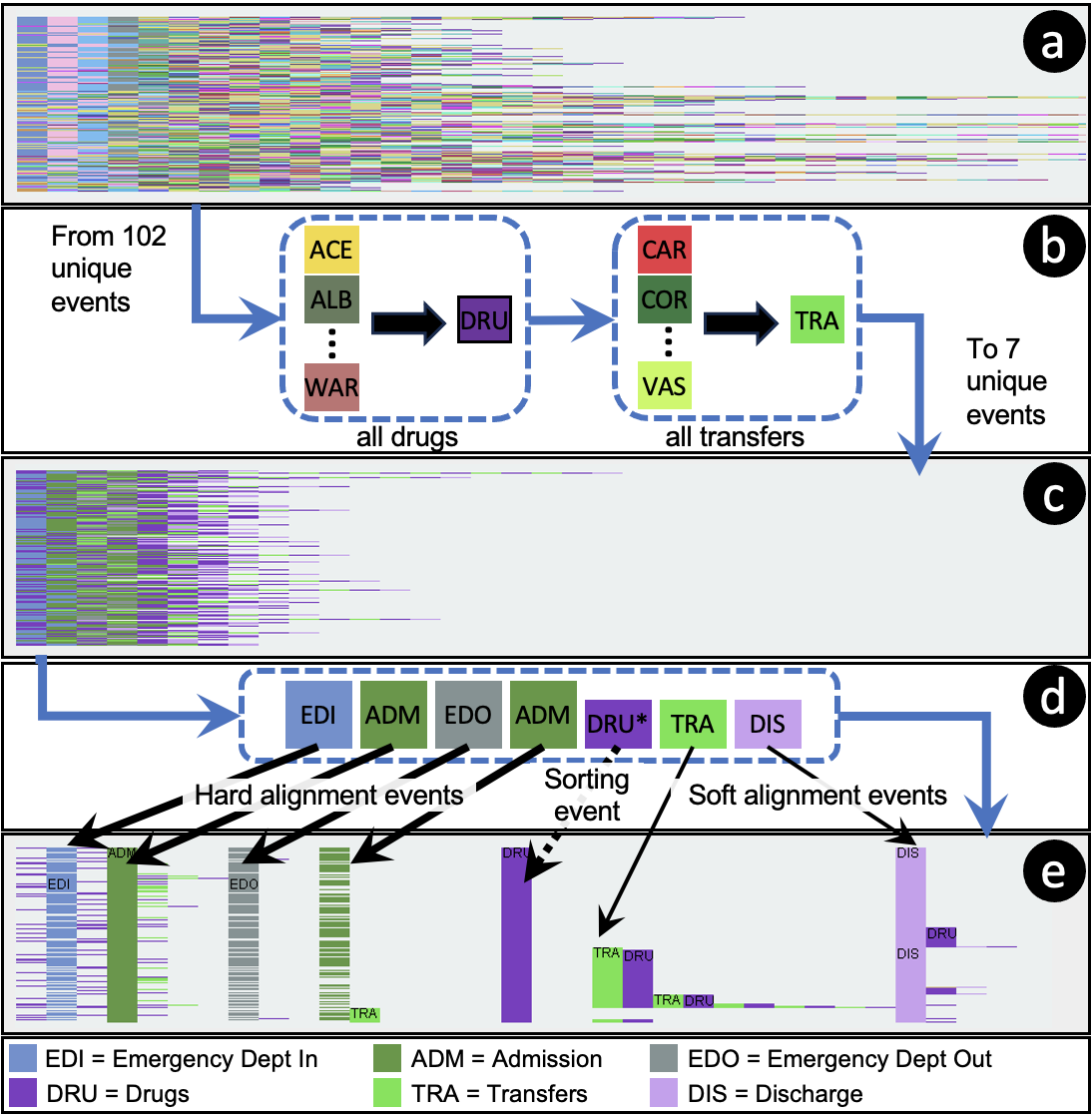}
   \caption{MIMIC-IV. (a) Starting overview. (b) Substitution and aggregation. (c) Intermediate overview. (d) Alignment and sorting. (e) Final overview. }
   \label{fig:mimic}
\end{figure}
\subsection{MIMIC-IV}\label{sub:caseStudyMIMIC}

The MIMIC-IV database \cite{johnson2023MIMICIV} contains data from patient admissions at a tertiary medical center, organized into relational tables that include demographic information and timestamped records of all clinical events from admission to discharge. The database includes numerous patient attributes, such as medications and laboratory measurements. For this case study, an individual sequence was defined to represent all timestamped events for a single admission, obtaining the data from tables including \texttt{admissions}, \texttt{diagnoses\_icd}, \texttt{patients}, \texttt{prescriptions}, \texttt{services}, and \texttt{transfers}. As we were interested in studying the management of patients with Hypertension, Chronic Kidney Disease, and Diabetes, we filtered by the relevant clinical codes. As a result, our dataset had 1,776 patients with at least two of these conditions. 

The case study aims to demonstrate the value that data transformations \textbf{(\transform)} in Sequen-C have for identifying patterns within complex datasets. \autoref{fig:mimic}(a) shows an initial cluster, including 102 types of events with sequences of up to 153 events in length. The diversity and length of these sequences initially complicate the identification of any discernible patterns within the cluster. However, \autoref{fig:mimic}(c) shows how substitution and aggregation of event types into user-defined event types significantly reduces the complexity, shortening the sequences and decreasing the number of event types to 7. In this case, all drugs have even replaced and aggregated by the more generic event called \textit{DRU} (drugs), and similarly, all transfers between clinics by the event \textit{TRA} (transfers). This step begins to reveal underlying patterns, but the visualization can still be improved by aligning and sorting the sequences. \autoref{fig:mimic}(e) shows the final overview after applying the transformation in \autoref{fig:mimic}(d). The resulting overview facilitates understanding the proportion of transfers in the dataset and how those relate to changes in medications before the final discharge event \textit{DIS}.

\section{Discussion, Limitations and Future Work}\label{section:discussion}

\textbf{Discussion}. The case studies and evaluation confirm our approach's effectiveness in supporting pattern discovery and data exploration. Domain experts emphasized its flexibility and efficiency in identifying outliers and bottlenecks, while novice users noted a steep learning curve and usability issues. These differences likely stem from varying expertise levels and hardware limitations during the evaluation.

The high evaluation accuracy demonstrates that Sequen-C and EventBox meet the design requirements, even for novice data analysts. This indicates that users with no prior experience can quickly grasp key concepts (e.g., EventBox encoding, event sequences, clustering, and alignment) and effectively use the visualizations to extract meaningful insights from the data.

Domain experts highlighted the prevalence of data quality issues, recommending the integration of strategies to detect and manage them. In contrast, novice analysts prioritized system feedback, specifically noting the lack of explicit alerts for missing or duplicate data. These differences likely reflect the participants' distinct backgrounds and priorities when interacting with Sequen-C and EventBox.

Although we did not conduct a direct comparison between Sequen-C with and without EventBox, our experience with both systems suggests that EventBox significantly enhances the analytical capabilities of Sequen-C. EventBox enables complex analyses that Sequen-C alone cannot support, while Sequen-C’s features, in turn, facilitate the effective use of EventBox. This synergy substantially reduces the time required for conducting intricate analyses, demonstrating the complementary strengths of the two systems.

\textbf{Limitations}. The approach faces limitations in scalability and analytic provenance. Large, complex datasets can impact performance and cause cognitive overload. While current strategies such as filtering, aggregation, and heatmaps help reduce visual clutter, further improvements, such as true heatmaps and automatic feature selection, are needed. Additionally, the exploratory nature of the system may affect reproducibility. Capturing analytic provenance \cite{xu2020Survey} could improve the consistency and interpretability of findings.

\textbf{Future work}. Combining the existing statistical reports with analytic provenance features and large language models could prove valuable for automatic storytelling. Using EventBox's multivariate analysis capabilities and the statistical report as a starting point also offers the possibility of enabling the construction of interactive predictive models. 
Lastly, performing an unstructured evaluation of EventBox, outside Sequen-C, would provide a clearer assessment of its standalone utility and generalizability.

\section{Conclusion}\label{section:conclusions}

This paper presents EventBox, a novel data representation and visual encoding for the analysis of multivariate event sequences. To demonstrate its utility, we integrated it into Sequen-C, a visual analytics system that builds dynamic, multilevel overviews of event data. EventBox enables simultaneous exploration of temporal and attribute-based patterns, supported by interactive transformations and quantitative overlays to reduce visual bias. An automatically generated statistical report complements the visual analysis by revealing significant attribute interactions. Evaluation results and case studies confirm the effectiveness of our approach, particularly in conveying relationships between variables, as reflected in high Insight and Essence scores in the ICE-T framework.

\clearpage

\acknowledgments{We gratefully acknowledge the contribution of the Sheffield Teaching Hospital NHS Trust for providing the Antenatal Care outpatient clinic dataset, and the NHS Trusts in the Yorkshire and the Humber region, which provided the original data to the CUREd Research Database. This report is independent research funded by the National Institute for Health and Care Research, Yorkshire and Humber Applied Research Collaborations NIHR200166. The views expressed in this publication are those of the author(s) and not necessarily those of the NHS, the National Institute for Health and Care Research or the Department of Health and Social Care.
This study is also supported by the European Union's H2020 programme under grant agreements CompBioMed, CompBioMed2, and Sano (Nos. 675451, 823712, and 857533).
The authors also wish to thank the participants in the \sequenCP user evaluation.}

\bibliographystyle{Bibliography/abbrv-doi-hyperref-narrow}
\bibliography{Bibliography/mainManuscript.bib}

\end{document}